\RequirePackage{fix-cm}

\documentclass[smallextended]{svjour3}   

\usepackage{xcolor}
\usepackage{graphicx}
\usepackage{mathptmx}
\usepackage{subfigure}
\usepackage{color}
\usepackage{amsmath}
\usepackage{lipsum}
\usepackage{amssymb}
\usepackage{epstopdf}
\usepackage[linktocpage,colorlinks=true,linkcolor=blue,citecolor=blue,breaklinks=true]{hyperref}
\usepackage{verbatim}
\usepackage{breakcites}
\usepackage[version=3]{mhchem}

\newcommand{\be}{\begin{equation}}
\newcommand{\ee}{\end{equation}}

\begin{document}


\title{ \textcolor{black}{Analytical Survival Analysis of the Ornstein-Uhlenbeck Process}}

\author{L. T. Giorgini \and W. Moon   \and J. S. Wettlaufer 
}

\institute{L. T. Giorgini \at Nordita, Royal Institute of Technology and Stockholm University, Stockholm 106 91, Sweden\\
  \email{ludovico.giorgini@su.se}\\
		\and W. Moon \at Department of Mathematics, Stockholm University 106 91 Stockholm, Sweden\\ 
			Nordita, Royal Institute of Technology and Stockholm University, Stockholm 106 91, Sweden\\
			  \email{woosok.moon@su.se}\\
                       \and
            J. S. Wettlaufer \at
             Yale University, New Haven, USA \\
                         Nordita, Royal Institute of Technology and Stockholm University, SE-10691 Stockholm, Sweden \\
               \email{john.wettlaufer@yale.edu}
}

\date{Received: date / Accepted: date}

\maketitle

\begin{abstract}
We use asymptotic methods from the theory of differential equations to obtain an analytical expression for the survival probability of an Ornstein-Uhlenbeck process with a potential defined over a broad domain. 
We form a uniformly continuous analytical solution covering the entire domain by asymptotically matching approximate solutions in an interior region, centered around the origin, to those in boundary layers, near the lateral boundaries of the domain.  The analytic solution agrees extremely well with the numerical solution and 
takes into account the non-negligible leakage of probability
that occurs at short times when the stochastic process begins close to one of the boundaries. Given the range of applications of Ornstein-Uhlenbeck processes, the analytic solution is of broad relevance across many fields of natural and engineering science.
\end{abstract}


\section{Introduction}

The generalized Ornstein-Uhlenbeck (OU) model describes a stochastic process  with at least one equilibrium point. 
It provides a framework for a wide range of physical, biological and social systems, wherein stabilization is viewed in terms of a potential minimum, characterized by a negative Lyapunov exponent, and high-frequency fluctuations are interpreted in terms of 
specific noise forcing.  For example, an OU process is used to study neuronal activity \cite{Ricciardi1979} and the time-evolution of trait values towards their evolutionary optima \cite{OMeara2014}.  In a clinical setting the health of the hepatic dynamic equilibrium is fit to an OU process using maximum likelihood estimation.  Stochastic volatility, crucial for deducing stock returns or option pricing, is treated in terms of an OU process \cite{Schobel1999}, as are the noise spectra of climate observations  \cite{Hasselmann1976, Moon2017, MAW2018}.

A particular stochastic model is commonly studied in terms of the {\em survival probability}, which is associated with \textcolor{black}{the probability of one or more events occurring, or for the system to reach a defined threshold for the first time starting from a given initial position}.  Due to the generality of the question it addresses, survival analysis has been widely used in science and engineering.  Examples include Feshbach resonances and 
the quantum Zeno effect (for example \cite{Feshbach,Zeno1,Zeno2}), engineering reliability analysis \cite{Zacks2012},  financial risk management \cite{McNeil2015},  and event history analysis
in sociology \cite{Blossfeld2014}. Moreover, in the specific case of an OU process survival analyses from neuroscience \cite{Tuckwell1988} and epidemiology \cite{Mode2000, Madec2004} to quantitative finance \cite{Leblanc1998, Linetsky2004, Jeanblanc2000} \textcolor{black}{and extreme value statistics of correlated random variables \cite{Majumdar2020}} demonstrate the ubiquity of the approach. 
\textcolor{black}{Our survival analysis is broadly relevant to all systems that can be described by an OU process. For example, it can be shown \cite{Gautie2019} that the survival probability of Brownian motion with an absorbing boundary that moves in time $t$ as $\propto \sqrt{t}$ can be recast as an OU process with a fixed absorbing boundary using Lamperti's (or Doob's) transformation.}  

\textcolor{black}{Let $\{X(t),\, t \geq0\}$ be an OU process starting at $x$.} The time it takes for the state of a system to encounter a threshold $\textcolor{black}{X(t)=\beta}$ for the first time is variously called the first hitting time or first passage time, whose probability distribution function $\textcolor{black}{\zeta(t,x)}$, is defined as
\begin{equation}
\textcolor{black}{\zeta(t,x)=\frac{\partial}{\partial t}\textrm{Prob}\{T \leq t\}}, 
\end{equation}
\textcolor{black}{where} 
\begin{equation}
\textcolor{black}{T\equiv\textrm{inf}\{t\,:\,X(t)>\beta|X(0)=x\}.}
\end{equation}
The time integral of the probability distribution of the first passage time is the survival probability (\textcolor{black}{see for example \cite{Bray2013, Redner2001, Alili2005, Rich} for reviews}). 

Despite the broad applicability of the survival probability, \textcolor{black}{a simple accurate analytical expression has been lacking, which is the primary motivation of our work}.
The exact mathematical form is constructed using the Laplace transform and its inverse, 
which results in a series expansion of special functions {\cite{Darling1953}}.  However, its complexity confounds practical implementation.  
In particular, when the initial data are close to the boundary of the confining potential \cite{Dolgoarshinnykh2006}, or when the boundary itself is near the equilibrium point \cite{Ditlevsen2005}, one must retain a considerable number of terms in the expansion.  

\begin{figure}
	\centering
	\includegraphics[width=0.9\textwidth]{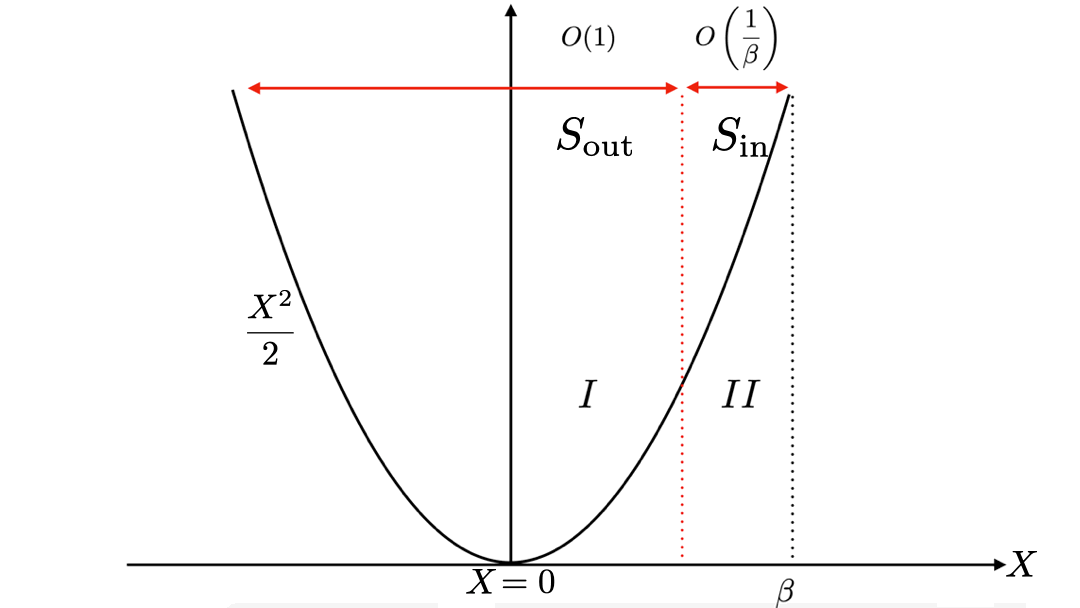}
	\caption{\footnotesize{Schematic of the problem under study; what is the probability that a particle  reaches the edge of the potential well?  We divide the potential into an ``outer'' region $I$ and a boundary layer $II$, or the ``inner'' region, of the potential, in each of which the asymptotically dominant solutions are determined and then matched.}}
	\vspace{-0.15 in}
	\label{F1}
\end{figure} 

Our treatment of survival probability is shown in the schematic potential of Fig. \ref{F1}, which contains a particle governed by a
one-dimensional Ornstein-Uhlenbeck process through an overdamped autonomous Langevin equation given by
\begin{equation}
\textcolor{black}{\dot{X}(t)=-aX(t)+\sqrt{2 b}\xi(t)},
\label{langD}
\end{equation}
where $a$ and $b$ are positive constants and $\xi(t)$ is Gaussian white noise with  
a zero mean and $\langle \xi(t)\xi(s)\rangle=\delta(t-s)$.  

The survival probability commonly characterizes the anomalous or abnormal behavior of a system.  This motivates our consideration of a threshold state ($\textcolor{black}{X=\beta}$) far from the equilibrium state ($\textcolor{black}{X(t)=0}$), so that it is rare that the system will reach the threshold.  

\textcolor{black}{The next section is devoted to our overall approach. In \S 2.1 we summarize our method. In \S 2.2 we briefly review previous results on the survival probability of an OU process that will be used in \S 2.3 to derive a simple analytical expression that reproduces the survival probability for large values of the domain boundaries and when the initial data are close to the boundary. In \S 2.4 we compare the analytical and numerical results.}


\section{Using Matched Asymptotic Methods to Determine the Survival Probability}

\subsection{Summary of Analytical Method}

Our approach is as follows.  As shown in Fig. \ref{F1} we divide the domain into two regions \cite{footnote}:
a broad $O(1)$ region ($I$) containing the minimum of the potential, $\textcolor{black}{X=0}$, and a narrow $O\left({1}/{\beta}\right)$ boundary layer near $\textcolor{black}{X=\beta}$.  We solve the limiting differential equations in these regions, from which we develop a uniform composite solution for the probability density of the survival probability using asymptotic matching, which is a highly accurate general analytical method \cite{Bender2013}, recently used to obtain the solution to the related problem of stochastic resonance \cite{Moon2020}.

\subsection{\textcolor{black}{Previous Results on the Survival Probability of an OU Process}}

The canonical approach of finding the survival probability of this OU process 
is to determine the probability distribution for the first hitting time 
in Laplace space
\begin{equation}
\textcolor{black}{\tilde{\zeta}(\lambda,x)=\int_{0}^\infty e^{-\lambda t}\zeta(t,x) dt,}
\end{equation} 
which leads to the following analytical expression
\cite{Siegert1951, Srinivasan2013, Ricciardi2013},
\begin{equation}
\tilde{\zeta}(\lambda,x)= \, e^{\frac{a(x^2-\beta^2)}{4b}}\frac{D_{-\lambda / a}(-x \sqrt{a/b})}{D_{-\lambda / a}(-\beta \sqrt{a/b})},
\label{ILzeta}
\end{equation}
where $D_{\lambda}(z)$ is the parabolic cylinder function, 
$\beta$ is the boundary position and $x$ is the initial position.

This equation can be written as a spectral decomposition \cite{Ricciardi1988}, 
wherein the countable number of eigenvalues correspond to the zeros of 
the denominator of Eq. \eqref{ILzeta}. The final expression 
for the probability distribution of the first hitting time can be obtained by inverting 
each term of this spectral decomposition, which can be written 
as a weighted sum of exponential functions as
\begin{equation}\begin{split}
 \zeta(t,x)=\sum_{p=0}^\infty e^{\lambda_p(\beta) t}\frac{\Phi(\lambda_p(\beta) / a,x \sqrt{a/b})}{\Phi'(\lambda_p(\beta) / a,\beta \sqrt{a/b})},
 \label{Survival_ricc}
\end{split}\end{equation}
where $\lambda_p(\beta)$ is the solution of 
\begin{equation}
\Phi(\lambda_p(\beta) / a,\beta \sqrt{a/b})=0,
\end{equation}
and $\Phi$ and $\Phi'$ are defined in \cite{Ricciardi1988} 
in terms of Kummer and digamma functions.

For large values of $t$ and $\beta-x$, Eq. (\ref{Survival_ricc}) simplifies considerably. 
In fact, because all the eigenvalues, $\lambda_p(\beta)$, are negative, 
only that with the smallest magnitude will contribute significantly as $t \to \infty$. 
Moreover, for large values of $\beta-x$, Eq. (\ref{Survival_ricc}) becomes
\begin{equation}\begin{split}
\lim_{\beta-x\to \infty}\zeta(t,x) &=\frac{1}{t_1(\beta;0)}\exp\left[-\frac{t}{t_1(\beta;0)}\right] \qquad \textrm{with}\\
t_1(\beta;0) &= \frac{1}{a\beta} \sqrt{\frac{2 \pi b}{a}}\exp\left[\frac{a\beta^2}{2b}\right],
\label{mean_time}
\end{split}\end{equation}
where $t_1(\beta;0)$ is the mean first passage time of the process from $X(t=0)=0$ to the boundary $\beta$ \cite{Nobile1985}.

Importantly, however, the asymptotic expression in Eq. \eqref{mean_time} is not valid when $x\approx\beta$. Indeed, as we show here, when the process starts in the neighborhood of $x=\beta$ there is a non-trivial leakage of probability.  This leakage is not 
taken into account by Eq. \eqref{mean_time} and transpires very rapidly, on a time scale of order $1/\beta^2$.  Therefore, taking this approach requires that one calculate an enormous number of eigenvalues in Eq. \eqref{Survival_ricc}, which is computationally inefficient. Our approach avoids this problem.  

\subsection{Detailed Development of the Approach}

We derive an asymptotic expression for the survival probability, which is the time integral of the probability distribution of the first passage time,
in the large $\beta$ limit that is trivial to evaluate when $x\approx\beta$.  

The probability density, $\rho(y,t;x,s)$, of the OU process in Eq. \eqref{langD}
is described by the Kolmogorov forward (KFE) and backward (KBE) equations, the former of which is
\begin{align}
\label{eq_KFE}
\partial_t \rho(y,t;x,s)&=a~\partial_y[y(t)\rho(y,t;x,s)] 
\\ \nonumber &+b~\partial_{yy}\rho(y,t;x,s)\,\equiv\mathcal{L}_y\rho(y,t;x,s),
\end{align}
%
%
%
where the operator $\mathcal{L}_y$ is the generator of the OU process viz., $[\mathcal{L}_y f](y,t)=a~\partial_y [y f(y,t)]+b~\partial_{yy}f(y,t)$. The KBE follows by replacing $\mathcal{L}_y$ with its adjoint, $\mathcal{L}_x^*$, defined as $[\mathcal{L}_x^* f](x,s)=-ax~\partial_x  f(x,s)+b~\partial_{xx}f(x,s)$.  
  
The KFE gives the evolution of the probability density of the process when the {\em initial} position and time, ($x,s$), 
are known, while the KBE treats the evolution when the {\em final} position and time, ($y,t$), are known.
Both equations have initial condition
\begin{equation}
\rho(y,u;x,u)=\delta(y-x),
 \label{ic}
\end{equation}
and boundary conditions
\begin{equation}
\rho(\pm \infty,t;x,s)=0 \qquad \textrm{and} \qquad \rho(y,t;\pm \infty,s)=0
 \label{bc_F}
\end{equation}
for the KFE and the KBE respectively. 
%

The survival probability within the interval $(\alpha,\beta)$ is defined in terms of the KBE density, $\rho_{KBE}$, as
\begin{equation}
S(t;x,s)=\int_{\alpha}^{\beta}\rho_{KBE}(y,t;x,s)dy,
\label{S_KBE}
\end{equation}
which satisfies 
\begin{equation}\begin{split}
-\partial_s S(t;x,s)=-a~x(s)\partial_x S(t;x,s)+b~\partial_{xx} S(t;x,s),
\label{Survival}
\end{split}\end{equation}
with initial condition
\begin{equation}
S(t=s;x,s)=\Theta(x-\alpha)\Theta(\beta-x),
\label{surv_cond_t}
\end{equation}
and boundary conditions
\begin{equation}
S(t;x=\alpha,s)=S(t;x=\beta,s)=0  \;\;\; \forall \;s \leq t,\label{surv_cond_x} 
\end{equation}
where $\Theta(\cdot)$ is the Heaviside theta function.

Let $P_i^\alpha (P_i^\beta)$ be the probability of hitting $x=\alpha (\beta)$ for the first time after $i$ time steps and let $S^\alpha (S^\beta)$ be the survival probability with an absorbing boundary at $x=\alpha (\beta)$.  Hence, if we discretize 
the stochastic process, the survival probability after $n$ time steps is
\begin{equation}\begin{split}
S_n&=\prod_{i=1}^n(1-P_i^\alpha-P_i^\beta)=\prod_{i=1}^n[(1-P_i^\alpha)(1-P_i^\beta)-P_i^\alpha P_i^\beta]\\&\simeq \prod_{i=1}^n(1-P_i^\alpha)\prod_{i=1}^n(1-P_i^\beta)=S_n^\alpha S_n^\beta.
\label{discret_S}
\end{split}\end{equation}  
In the second line of Eq. \eqref{discret_S}, we have neglected the term $P_i^\alpha P_i^\beta$ 
when $|\alpha|\gg 1$ and $|\beta| \gg 1$, allowing us to write the survival probability in the interval $(\alpha, \beta)$ 
as the product of the two survival probabilities in the two intervals 
$(-\infty, \beta)$ and $(\alpha, \infty)$ with $\alpha<\beta$.  From this point we will only consider the survival probability 
in the interval $(-\infty, \beta)$, and note that the derivation for the interval $(\alpha, \infty)$ is in straightforward analogy.  

We now rewrite Eq. \eqref{Survival} in a rescaled form, 
\begin{equation}\begin{split}
\partial_t S(x,t)=-x(t)\partial_x S(x,t)+\partial_{xx} S(x,t),
\label{SurvivalS}
\end{split}\end{equation}
with the new variables, 
\begin{equation}
\label{scales}
x \to x\sigma, \;\;\; \beta \to \beta\sigma ~~\textrm{and}~~ s \to -\frac{t}{a},
\end{equation}
in which the spatial coordinate is expressed in terms of the standard deviation $\sigma=\sqrt{b/a}$ of a stationary OU process obtained from the solution of Eq. \eqref{eq_KFE} in the limit $t \to\infty$.  
%
The new initial and boundary conditions are
\begin{equation}
\begin{cases}
S(x,t=0)=\Theta(\beta-x), \\
 S(x=\beta,t)=0,\\ S(x=-\infty,t)=1.
\end{cases}
\label{conditions}
\end{equation}
We solve the limiting differential equations within the two regions, from which we construct an approximate uniform solution by asymptotic matching.
We denote $S_{\textrm{out}}(x,t)$ and $S_{\textrm{in}}(x,t)$ the solutions in region $I$ and $II$  respectively.    

\underline{Region $I$}.  The outer solution is obtained by imposing $\beta-x \gg 1$.  In this limit the probability distribution of the first hitting time in Eq. \eqref{mean_time} is valid, and its time integral gives the survival probability as 
\begin{equation}
S_{\textrm{out}}(t){\simeq}\,\textrm{e}^{-\left(\frac{\beta}{\sqrt{2 \pi}} \textrm{e}^{-\frac{\beta^2}{2} } \right) t} ,
\label{Sinner}
\end{equation}
where the rescaled $\beta$ and $t$ of Eq. \eqref{scales} have been used.

\underline{Region $II$}. In the boundary layer, or the inner region, we have $x \sim \beta$, where the approximation of Region $I$ is no longer valid. We let $\epsilon \equiv 1/\beta\ll 1$ and introduce the following stretched coordinates, 
\begin{equation}
\eta=\frac{x-\frac{1}{\epsilon}}{\epsilon} ~~\textrm{and}~~  \theta=\frac{t}{\epsilon^2}, 
\end{equation}
that we use to rewrite Eq. (\ref{SurvivalS}) as
\begin{align}
\frac{1}{\epsilon^2}\partial_\theta S_{\textrm{in}}(\eta,\theta)&=-\left[\frac{1}{\epsilon^2}+\eta(\theta)\right]\partial_\eta S_{\textrm{in}}(\eta,\theta)
\nonumber \\
&+\frac{1}{\epsilon^2}\partial_{\eta\eta} S_{\textrm{in}}(\eta,\theta), 
\label{SurvivalStr}
\end{align}
which at leading-order becomes
\begin{equation}
\partial_\theta S_{\textrm{in}}(\eta,\theta)=-\partial_\eta S_{\textrm{in}}(\eta,\theta)+\partial_{\eta\eta} S_{\textrm{in}}(\eta,\theta).
\label{SurvivalLeading}
\end{equation}
Clearly, Eq. \eqref{SurvivalLeading} is a diffusion equation for $S_{\textrm{in}}(\eta,\theta)$ along the characteristics
\begin{equation}
\frac{d\eta}{d\theta}=1 ~~\textrm{and}~~ \frac{d\rho}{d\theta}=1, 
\end{equation}
which we can then write as
\begin{equation}
\partial_\rho S_{\textrm{in}}(\mu,\rho)=\partial_{\mu\mu} S_{\textrm{in}}(\mu,\rho),
\label{s_diff}
\end{equation}
wherein $\mu\equiv\eta-\theta$ and $\rho\equiv\theta$, so that the boundary condition  becomes $S_{\textrm{in}}(\mu=-\rho,\rho)=0$.

We solve Eq. \eqref{s_diff} by first finding its Green's function, $G(\rho,\mu;\nu)$, which satisfies
\begin{equation}
\partial_\rho G(\mu,\rho;\nu)-\partial_{\mu\mu} G(\mu,\rho;\nu)=\delta(\mu-\nu)\delta(\rho).
\end{equation}
This Green's function is associated with the probability density $\rho_{KBE}$, satisfying the KBE, and hence the survival probability through  Eq. \eqref{S_KBE}.  Hence, this density satisfies the following conditions;
\begin{equation}
\begin{cases}
\rho_{KBE}(\mu,\rho=0;\nu)=\delta(\mu-\nu), \\
 \rho_{KBE}(\mu=0,\rho;\nu)=\rho_{KBE}(\mu=-\infty,\rho;\nu)=0.
\end{cases}
\label{conditionsKBE}
\end{equation}
The Green's function is (e.g., \cite{Duffy2015})
\begin{equation}\begin{split}
G(\rho,\mu;\nu)=&\frac{1}{\sqrt{4\pi\,\rho}}\bigg(\exp\left[-\frac{(\nu-\mu)^2}{4\,\rho}\right]-\\&-\exp\left[-\frac{(\nu+\mu)^2}{4\,\rho}-\nu\right]\bigg), 
\end{split}\end{equation}
which, as noted above, now allows us to write the solution of Eq. (\ref{s_diff}) as
\begin{equation}\begin{split}
S_{\textrm{in}}(\mu,\rho)&=\int_{-\infty}^0d\nu\int_{-\infty}^\infty d\phi \,G(\mu,\rho;\phi)\delta(\phi-\nu)=\\&=\int_{-\infty}^0 d\nu \,G(\mu,\rho;\nu).
\end{split}\end{equation}
Upon integration and reversion to the original 
variable $t$ and to the stretched coordinate $\eta$, we find
\begin{equation}\begin{split}
S_{\textrm{in}}(\eta, t){\simeq}&\frac{K}{2}\textrm{e}^{\eta}\bigg(-\textrm{erfc}\bigg[\frac{1}{2}\sqrt{\frac{1}{t}}(-t-\eta)\bigg]+\\&+\textrm{e}^{- \eta}\textrm{erfc}\bigg[\frac{1}{2}\sqrt{\frac{1}{t}}(-t+\eta)\bigg]\bigg)\textcolor{black}{\equiv K \,S_{\textrm{in}}'(\eta, t)}.
\label{Souter1}
\end{split}\end{equation}

We determine the constant $K$ by requiring the outer limit of the inner solution to equal the outer solution; 
\begin{equation}\begin{split}
&K=\lim_{\eta \to -\infty}S_{\textrm{in}}(\eta,t)=S_{\textrm{out}}(t){\simeq}\textrm{e}^{-\left(\frac{\beta}{\sqrt{2 \pi}} \textrm{e}^{-\frac{\beta^2}{2} } \right) t} .
\label{match2}
\end{split}\end{equation}
Therefore, the uniformly valid approximate composite analytical solution for the survival probability is 
\begin{equation}\begin{split}
S(x,t)&=S_{\textrm{in}}(x,t)+S_{\textrm{out}}(t)-K \\&{\simeq}\frac{1}{2}\textrm{e}^{-\left(\frac{\beta}{\sqrt{2 \pi}} \textrm{e}^{-\frac{\beta^2}{2} } \right) t}
\textrm{e}^{\beta(x-\beta)} \\ & \bigg(\textcolor{black}{-~}\textrm{erfc}\bigg[\frac{1}{2}\sqrt{\frac{1}{t}}(-t-\beta(x-\beta))\bigg]+\\&+\textrm{e}^{- \beta(x-\beta)}\textrm{erfc}\bigg[\frac{1}{2}\sqrt{\frac{1}{t}}(-t+\beta(x-\beta))\bigg]\bigg)=\\&\equiv S\textcolor{black}{'}_{\textrm{in}}(x,t)S_{\textrm{out}}(t),
\label{Souter}
\end{split}\end{equation}
written in terms of the original spatial coordinate $x$ rather that the stretched coordinate $\eta$.
\subsection{\textcolor{black}{Comparing Analytical and Numerical Results}}

\begin{figure}
	\centering
	\includegraphics[width=0.75\textwidth]{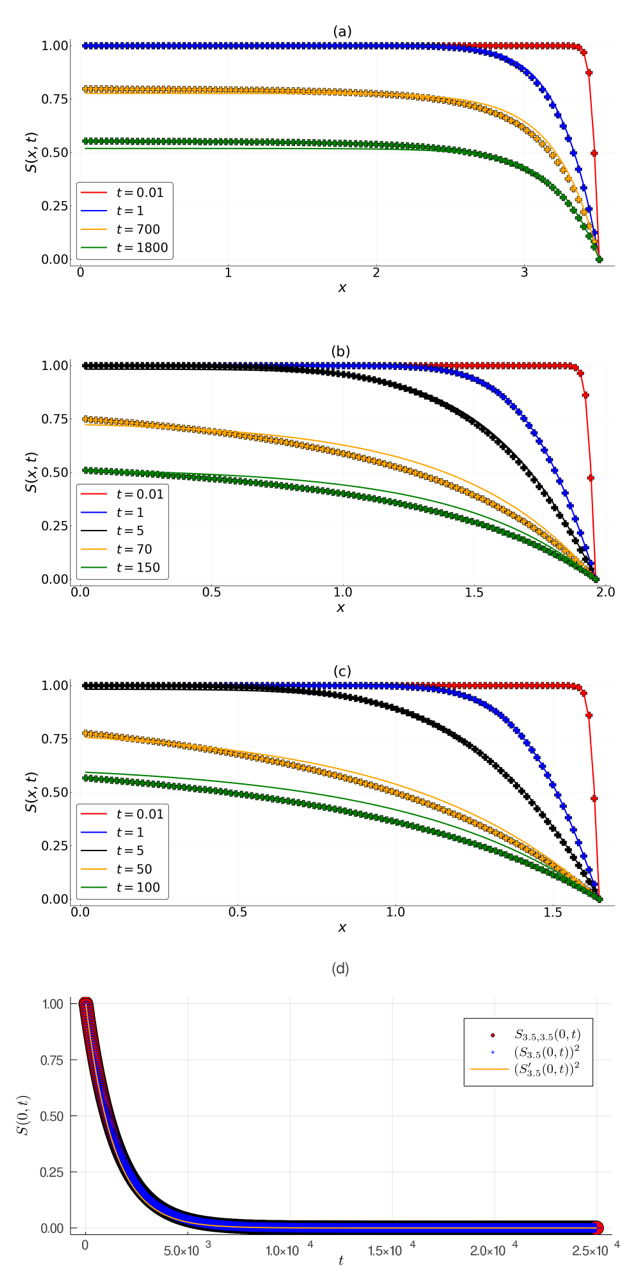}
	\caption{\footnotesize{
	(a)-(c) Plot of Eq. (\ref{Souter}), $S(x,t)$,  versus $x \in [0,\beta]$ for different values of $t$ (solid lines) compared to the numerical solution of Eq. (\ref{SurvivalS}) (crosses). We use three different values of the boundary position $\beta$; (a) 3.5 (b) 1.96 and (c) 1.645,  corresponding to the probability of finding the system below $x=\beta$ as $t \to \infty$ and without absorbing boundaries of (a) $99.95\%$, (b) $99\%$, and (c) $90\%$ respectively.  (d) The numerical solution of Eq. (\ref{SurvivalS}) with two boundaries (red circles), $S_{\alpha,\beta}(x,t)$, $S_{\alpha}(x,t) S_{\beta}(x,t)$ (blue crosses), and  the approximate analytical solution in Eq. (\ref{surv_fin})$, S'_{\alpha}(x,t) S'_{\beta}(x,t)$.  Note that $S_{\alpha,\beta}(x,t)$ overlaps exactly with $S_{\alpha}(x,t)S_{\beta}(x,t)$. Here $\alpha=\beta=3.5$ and $x=0$.}}
	\vspace{-0.17 in}
	\label{F2}
\end{figure} 

We compare Eq. (\ref{Souter}) with the numerical solution of Eq. (\ref{SurvivalS}) for different values of the parameter $\beta$
in Fig. (\ref{F2}), and find excellent agreement even when $\beta$ is {\em not} asymptotically large.  Indeed, the smallest value of the distance to the boundary is $\beta=1.645 \sigma$, corresponding to a $90\%$ probability that an unbounded stationary OU process remains below the boundary.  

When $t\gg1/\beta^2$ the left hand side of Eq. (\ref{SurvivalLeading}) is negligible and  Eq. (\ref{Souter}) becomes $S_{\textrm{out}}(t)[1-e^{\beta(x-\beta)}]$ and depends on time solely through $S_{\textrm{out}}(t)$, which is the prefactor of $S_{\textrm{in}}(x,t)$.  Thus, 
 depending on $x$ the rate of the decrease in the survival probability is controlled by $S_{\textrm{in}}(x,t)$, decaying more rapidly near the boundary for early times.   
 
 The accuracy of the asymptotic solutions over a wide range of the $\beta$ facilitates simple and wide ranging applications.  For example,  determining the input parameters of a leaky integrate-and-fire (LIF) neural model (see e.g., \cite{Lansky2008} and refs therein)
 based on experimentally observable interspike intervals \cite{Ditlevsen2005}.  Of particular contemporary relevance is 
 deducing the ``critical community size'' in disease epidemiology \cite{Dolgoarshinnykh2006}, or the population threshold below which infections do not persist.

Finally, we appeal to Eq. \eqref{discret_S} to form the asymptotic expression for the survival probability of an Ornstein-Uhlenbeck process with two absorbing boundaries as
\begin{equation}
S_{\alpha,\beta}(x,t) = S\textcolor{black}{'}^{\alpha}_{\textrm{in}}(x,t) S^{\alpha}_{\textrm{out}}(t)S\textcolor{black}{'}^{\beta}_{\textrm{in}}(x,t) S^{\beta}_{\textrm{out}}(t)\; \forall t>s.
\label{surv_fin}
\end{equation}
We note that only one of the boundary regions will contribute significantly viz., 
\begin{equation}\begin{split}
& S\textcolor{black}{'}_{\textrm{in}}^{\alpha,\beta}(x,t) =S\textcolor{black}{'}^{\alpha}_{\textrm{in}}(x,t)S\textcolor{black}{'}^{\beta}_{\textrm{in}}(x,t) = S\textcolor{black}{'}^{\gamma}_{\textrm{in}}(x,t), \\ &\gamma = \textrm{min}_{|h_\alpha|, |h_\beta|}(\alpha,\beta),
\label{semp}
\end{split}\end{equation}
where $h_{\alpha(\beta)}=x-\alpha(\beta)$.

In Fig. (\ref{F2})(d) we show that the numerical solution with two boundaries matches that obtained by considering the two boundaries separately and with the analytical solution in Eq. (\ref{surv_fin}).

\section{Conclusion}

We have obtained an asymptotic analytical solution for the survival probability of an Ornstein-Uhlenbeck process for large potentials. We divide the potential into two layers near the boundaries and a broad region between, which contains the origin, where we use the solution of Ricciardi \& Sato \cite{Ricciardi1988}.   The uniformly continuous solution is obtained by matching the two approximate solutions in the boundary layers with that in the broad region between.  The solution agrees extremely well with both the numerical solution and with the more restricted asymptotic expression for the survival probability known in literature. 
Importantly, our analysis remains valid even when the initial position of the stochastic process is close to one of the boundaries, and furthermore it takes into account the non-negligible leakage of the probability early in the time evolution. 

Despite the analysis using the assumption of asymptotically large boundaries, we showed that it agrees well with the numerical solution, even when the boundary positions are the same order of magnitude as the standard deviation of the stationary probability distribution function of the OU process. We demonstrate consistency even when there is a $90 \%$ probability that the system without absorbing boundaries is found at a position less than $\beta$; when reaching the boundary can no longer  be considered a rare event. Therefore, our method can be easily generalized to more general (less restrictive) settings.  Finally, our compact analytical solution provides a computationally trivial framework for survival analysis of use across the broad spectrum of stochastic systems where the Ornstein-Uhlenbeck process arises. 

\section*{Acknowledgements} 
We gratefully acknowledge support from the Swedish Research Council Grant No. 638-2013-9243.

\begin{thebibliography}{10}
\bibitem{footnote}We note that in the parlance of the field, the boundary layer is the {\em inner region} and the remainder of the domain is the {\em outer region}, although in this case the latter is in the interior of the potential.
\end{thebibliography}
\end{document}